\documentstyle[epsf,prl,aps,twocolumn,floats]{revtex}
\begin{document}
\draft
\wideabs{
\title{CANTED ANTIFERROMAGNETIC PHASE IN A DOUBLE QUANTUM WELL IN A
TILTED QUANTIZING MAGNETIC FIELD}
\author{V.S.~Khrapai, E.V.~Deviatov, A.A.~Shashkin, V.T.~Dolgopolov}
\address{Institute of Solid State Physics, Chernogolovka, Moscow
District 142432, Russia}
\author{F.~Hastreiter, A.~Wixforth}
\address{Ludwig-Maximilians-Universit\"at, Geschwister-Scholl-Platz
1, D-80539 M\"unchen, Germany}
\author{K.L.~Campman, A.C.~Gossard}
\address{Materials Department and Center for Quantized Electronic
Structures, University of California, Santa Barbara, California
93106, USA}
\maketitle

\begin{abstract}
We investigate the double-layer electron system in a parabolic
quantum well at filling factor $\nu=2$ in a tilted magnetic field
using capacitance spectroscopy. The competition between two ground
states is found at the Zeeman splitting appreciably smaller than the
symmetric-antisymmetric splitting. Although at the transition point
the system breaks up into domains of the two competing states, the
activation energy turns out to be finite, signaling the occurrence of
a new insulator-insulator quantum phase transition. We interpret the
obtained results in terms of a predicted canted antiferromagnetic
phase.
\end{abstract}
}

Much interest in double-layer systems is aroused by the presence of
an additional degree of freedom which is associated with the third
dimension. In a double-layer system with symmetric electron density
distributions in a normal to the interface magnetic field at filling
factor $\nu=2$, the competition of different ground states is
expected that is controlled by the relation between the Coulomb
interaction energy, the spin splitting, and the
symmetric-antisymmetric splitting caused by interlayer tunneling. In
the simplest single-particle picture each Landau level has four
sublevels originating from the spin and subband splittings. With
increasing spin splitting a transition should occur from a spin
unpolarized ground state with anti-parallel spin orientations of
occupied sublevels to a ferromagnetic one with parallel spins when
the Zeeman energy $\mu gB$ is equal to the symmetric-antisymmetric
splitting $\Delta_{SAS}$. Experimentally, however, the clear
transition was observed at the Zeeman energy significantly smaller
than $\Delta_{SAS}$ \cite{boebinger,japan}, which points out the
importance of many-body effects for the transition. Recent
theoretical considerations
\cite{zheng,sarma,sarma1,demler,sarma2,mac} have revealed the crucial
role of electron-electron interaction for the spin structure of
double-layer electron systems. For the symmetric bilayer electron
system in a potential well that is stable to symmetry breaking
(so-called easy plane two-dimensional ferromagnet \cite{mac1}), in
addition to the consistent with found in experiment
\cite{boebinger,japan} shift of the phase transition point to smaller
magnetic fields, a new so-called canted antiferromagnetic phase
occurs between spin unpolarized and ferromagnetic state
\cite{zheng,sarma,sarma1,demler,sarma2}. It is shown that, due to
Coulomb repulsion of electrons, mixing the symmetric and
antisymmetric states with opposite spin directions forms a new ground
state which is a two-particle spin singlet. The transition point
between this spin singlet state and the ferromagnetic state in which
the spins in both layers point in the direction of the applied
magnetic field is defined by the relation

\begin{equation}
\mu gB\approx\frac{\Delta_{SAS}^2}{E_c} \label{eq1}\end{equation}
with $E_c$ denoting the Coulomb energy. Because normally $E_c>
\Delta_{SAS}$, the transition is expected at $\mu gB< \Delta_{SAS}$.
Near the transition the intralayer exchange interaction connects both
lowest states of the electron system and gives rise to the
intermediate canted antiferromagnetic phase that is characterized by
interlayer antiferromagnetic spin correlations in the two-dimensional
plane and is related to the zero-energy spin excitation mode
\cite{zheng,sarma,sarma1,demler,sarma2}. The appearance of this phase
signifies a new class of quantum phase transitions between insulators
with different spin structures.

The idea of a novel phase has been supported by the recent
experiments on inelastic light scattering in which transitions were
observed near $\nu=2$ between the ground states of a bilayer system
with different spin structures \cite{pel,pel1}. However, these states
so far have not been studied by transport measurement methods.

The existence of a canted antiferromagnetic phase is predicted for a
double layer with asymmetric electron density distributions as well;
moreover, external bias allows a continuous tuning of the $\nu=2$
state within a single gated sample \cite{sarma2}. As a result, in the
general case one has two transitions \cite{pel1} with increasing
bilayer asymmetry: ferromagnetic -- canted antiferromagnetic -- spin
unpolarized phase. For a sufficiently large ratio $\Delta_{SAS}/\mu
gB$, the canted antiferromagnetic phase becomes the ground state at
the symmetry/balance point, and the first transition disappears. A
further increase of the ratio leads to the disappearance of the
second transition. In a disordered system the interval between two
transition points is expected to have intrinsic structure and can
include different spin Bose glass phases \cite{demler,sarma2}.

Here, we employ a capacitance spectroscopy technique to study the
phase transition in the double-layer electron system in a parabolic
quantum well at filling factor $\nu=2$ at a tilted magnetic field.
The scenario of the observed transition gives strong evidence for a
new insulator-insulator quantum phase transition and supports the
formation of the recently predicted canted antiferromagnetic phase,
although for tilted magnetic fields a rigorous theory is not yet
available.

The samples are grown by molecular beam epitaxy on semi-insulating
GaAs substrate. The active layers form a 760~\AA\ wide parabolic
well. In the center of the well a 3 monolayer thick
Al$_x$Ga$_{1-x}$As ($x=0.3$) sheet is grown which serves as a tunnel
barrier between both parts on either side. The symmetrically doped
well is capped by 600~\AA\ AlGaAs and 40~\AA\ GaAs layers. The sample
has ohmic contacts (each of them is connected to both electron
systems in two parts of the well) and two gates on the crystal
surface with areas $120\times 120$ and $220\times 120$ $\mu$m$^2$.
The gate electrode enables us both to tune the carrier density in the
well, which is equal to $4.2\times 10^{11}$~cm$^{-2}$ at zero gate
bias, and measure the capacitance between the gate and the well. For
capacitance measurements we apply an ac voltage $V_{ac}=2.4$~mV at
frequencies $f$ in the range 3 to 600~Hz between the well and the
gate and measure both current components as a function of gate bias
$V_g$ in the temperature interval between 30~mK and 1.2~K at magnetic
fields of up to 14~T.

Our measurements are similar to magnetotransport measurements in
Corbino geometry: when disturbed the sample edge becomes
equipotential within the edge magnetoplasmon roundtrip time $\sim
L/\sigma_{xy}$ which is normally much shorter than the time of charge
redistribution normal to the edge $\sim C_0L^2/\sigma_{xx}$, where
$\sigma_{xx}$ and $\sigma_{xy}$ are the dissipative and Hall
conductivity, $C_0$ is the capacitance per unit area between gate and
quantum well, and $L$ is the characteristic sample dimension. At low
frequencies $f\ll \sigma_{xx}/C_0L^2$, the imaginary current
component reflects the thermodynamic density of states in a
double-layer system. In this limit, the, e.g., $\nu=2$ imaginary
current component minimum is accompanied by a peak in the active
current component which is proportional to $(fC_0)^2
\sigma_{xx}^{-1}$ and is used for measurements of the temperature
dependence of $\sigma_{xx}$. At high frequencies the minimum in the
imaginary current component should deepen as caused by in-plane
transport so that both current components tend to zero.

\begin{figure}[t]
\centerline{
\epsfxsize=\columnwidth
\epsffile{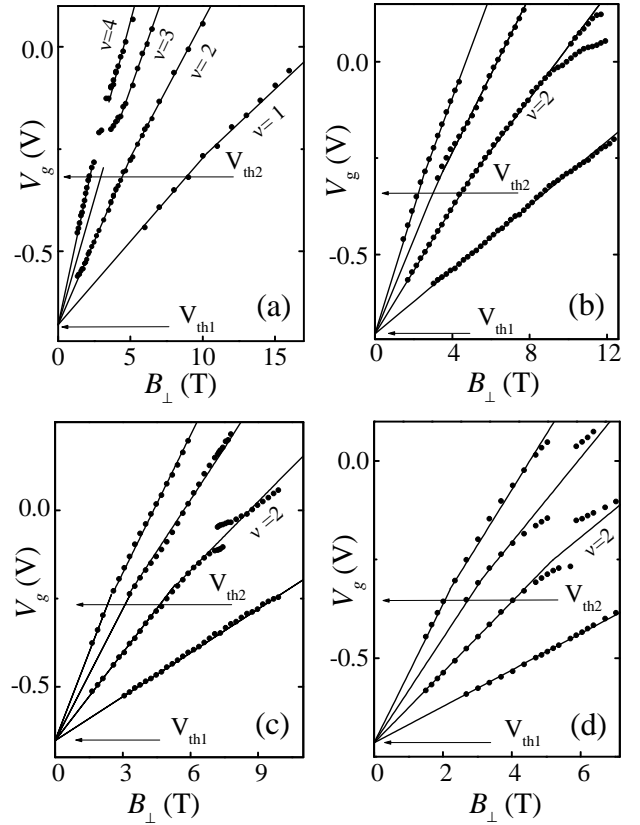}
}
\caption{Positions of the magnetocapacitance minima, or, of the
active current component maxima, at a temperature of 30~mK for
different tilt angles: (a) $\Theta=0^\circ$, (b) $\Theta=30^\circ$,
(c) $\Theta=45^\circ$, (d) $\Theta=60^\circ$.\label{exp}}
\end{figure}

The positions of the $\nu=2$ imaginary current component minimum (or,
active current component maximum) in the ($B_\perp,V_g$) plane are
shown in Fig.~\ref{exp} alongside with those at filling factors
$\nu=1,3,4$ for a normal and tilted magnetic fields. In the normal
magnetic field, at the gate voltages $V_{th1} <V_g< V_{th2}$, at
which one subband is filled with electrons in the back part of the
well with respect to the gate, the experimental points are placed
along a straight line with a slope defined by capacitance between the
gate and the bottom electron layer (Fig.~\ref{exp}a). Above
$V_{th2}$, where a second subband collects electrons in the front
part of the well, a minimum in the imaginary current component at
integer $\nu$ corresponds to a gap in the spectrum of the bilayer
electron system, and the slope is inversely proportional to the
capacitance between gate and top electron layer \cite{dol}. At a tilt
angle of 30$^\circ$, a splitting of the line indicating the position
of the $\nu=2$ minimum is observed close to the balance point
(Fig.~\ref{exp}b). Knowing $\Delta_{SAS}=1.3$~meV from far infrared
measurements and model calculations \cite{FIR} we can estimate from
Eq.~(\ref{eq1}) the Coulomb energy $E_c\approx 6$~meV at the
transition. This value is smaller than the energy $e^2/\varepsilon
l=15$~meV (where $l$ is the magnetic length) because of finite
extension of the electron wave functions in the $z$ direction. As
seen from Figs.~\ref{exp}c,\ref{exp}d, with increasing tilt angle
$\Theta$ the center of the splitting moves towards more negative gate
voltages. It is important that the occurrence of two distinct minima
at fixed magnetic field points to the competition between the two
ground states of our bilayer system.

\begin{figure}[t]
\centerline{
\epsfxsize=7.2cm
\epsffile{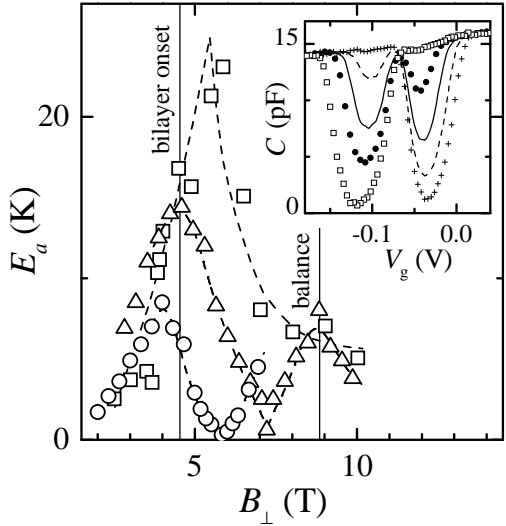}
}
\caption{Change of the activation energy at $\nu=2$ with magnetic
field for $\Theta=0^\circ$ (squares), $\Theta=45^\circ$ (triangles),
$\Theta=60^\circ$ (circles). The dashed lines are guides to the eye.
The data at $\Theta=30^\circ$ are not displayed to avoid
overcomplicating the figure. The inset shows the gate voltage
dependence of the imaginary current component for $\Theta=45^\circ$
at $f=23$~Hz at $T=30$~mK for different magnetic fields: 10.03
(squares), 10.19 (dots), 10.28 (solid line), 10.36 (dashed line), and
10.53~T (crosses).\label{EA}}
\end{figure}

Fig.~\ref{EA} represents the behaviour of the activation energy along
the $\nu=2$ line in Fig.~\ref{exp} at different tilt angles. For
$\Theta=0^\circ$ at $V_g> V_{th2}$, the activation energy passes
through a maximum and then monotonously decreases with increasing
magnetic field. At a tilted magnetic field there emerges a deep
minimum of the activation energy at the field corresponding to the
splitting point. We find that for all tilt angles the minimum
activation energy is finite (Fig.~\ref{EA}).

A set of the experimental traces near the splitting point for
$\Theta=45^\circ$ is displayed in the inset of Fig.~\ref{EA}. An
interplay is seen of two deep minima in the magnetocapacitance at
filling factors slightly above and slightly below $\nu=2$,
respectively, which correspond to maxima of the activation energy
with a minimum in between at exactly $\nu=2$. At the splitting point
both magnetocapacitance minima are observable simultaneously with
roughly equal amplitudes.

It is interesting to compare our experimental findings with results
of Ref.~\cite{japan} where a double layer system with higher mobility
and smaller $\Delta_{SAS}$ was investigated at $\nu=2$ at a normal
magnetic field. In both experiments a change of the ground state at
the balance point is reached by varying a tuning parameter: a total
electron density in Ref.~\cite{japan} and a tilt angle in our case.
In addition, we observe the coexistence of two ground states near the
transition point and find a finite value of the activation energy at
the transition, i.e., it is insulator-insulator transition.

To verify that our experimental results are beyond the
single-particle model we compare them with the single-particle
spectrum in a tilted magnetic field calculated in self-consistent
Hartree approximation (details of calculation will be published
elsewhere). In the calculation we do not take into account the spin
splitting (supposing small $g$ factor) as well as the exchange
energy.

\begin{figure}[t]
\centerline{
\epsfxsize=7.5cm
\epsffile{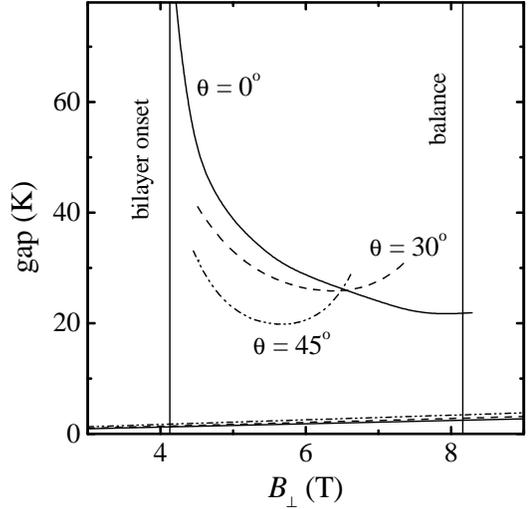}
}
\caption{Results of the $\nu=2$ gap calculation in self-consistent
Hartree approximation for $\Theta=0^\circ$, $\Theta=30^\circ$,
$\Theta=45^\circ$. Also shown is the corresponding Zeeman
splitting.\label{calc}}
\end{figure}

For this simplest case the calculated gap for filling factor $\nu=2$
also exhibits a minimum (Fig.~\ref{calc}). It is quite easy to
understand the physical origin of the minimum: on one hand, a
parallel component of the magnetic field leads to increasing the
subband energies because of narrowing the electron density
distribution in the $z$ direction. As a result, $\Delta_{SAS}$
increases a little. On the other hand, to form a gap in the spectrum
the tunneling between the layers should occur with conservation of
the in-plane momentum. Therefore, it is accompanied with a shift of
the center of the in-plane wave function by an amount $d_0\tan\Theta$
(where $d_0$ is the distance between the centers of mass for electron
density distributions in two lowest subbands) which enhances with
both deviation from the balance point due to the increase of $d_0$
and tilting the magnetic field. The increase of the effective
tunneling distance $\sqrt{d^2+d_0^2 \tan^2\Theta}$ (where $d$ is the
tunnel barrier width) results in decreasing the subband spacing. The
combination of these contributions, taking into account the known
behaviour of the $\nu=2$ gap in the normal magnetic field (see
Fig.~\ref{calc}), gives rise to the non-monotonous dependences of the
gap at tilted magnetic fields. From the calculation it follows that
the position of the minimum shifts to lower magnetic fields $B_\perp$
with increasing tilt angle, which is consistent with our experimental
finding. Nevertheless, the measured minimum activation energy is far
smaller than the calculated half-gap.

In principle, lower values of the measured activation energy might be
explained in the single-particle picture assuming that the quantum
level width is finite. However, we reject such an explanation for the
following reasons: first, at $\Theta=0^\circ$ at balance we do
observe an activation energy very close to half of $\Delta_{SAS}$ and
practically at the same magnetic field where a deep minimum in the
activation energy is observed at $\Theta=30^\circ$; second, when
sweeping the gate voltage through the splitting point the activation
energy shows two maxima with a non-zero minimum in between at
$\nu=2$, which indicates the coexistence of two ground states in the
form of domains at the critical point and is in contradiction to the
single-particle picture.

To apply the line-of-reasoning developed in a many-body lattice model
\cite{demler} to the case of a tilted magnetic field one has to
introduce two significant changes: (i) as one rung one should
consider two sites shifted on distance $d_0\tan\Theta$; (ii) the
value of $\Delta_{SAS}$ should be replaced by the subband spacing as
determined from self-consistent Hartree approximation. Then,
conclusion of Ref.~\cite{demler} about the existence of an
intermediate canted antiferromagnetic phase is expected to remain
valid. We note that the theory \cite{sarma,sarma1,demler,sarma2}
deals with a change of the ground state while experimentally we
measure the energy of a charge excitation with $k=\infty$ \cite{rem}.
Therefore, the comparison of theory with experiment is
straightforward only in the single-particle picture where the nearest
excited state is identical with the competing ground state. According
to Ref.~\cite{sarma1}, in the many-body problem peculiarities of the
activation energy are expected near the phase transition.

In our opinion, the observed deep minimum in activation energy at
tilted magnetic fields is a manifestation of transition from a spin
unpolarized state to a canted antiferromagnetic phase: at gate
voltages right above $V_{th2}$ the spin unpolarized state only can be
realized for filling factor $\nu=2$. The ferromagnetic and canted
antiferromagnetic phase should then be considered as possible states
near the balance point. At the transition point a finite activation
energy is found, which is not the case for the direct transition from
spin unpolarized to ferromagnetic state \cite{sarma1}. Hence, the
transition scenario forces us to recognize the phase around the
balance point as a canted antiferromagnetic phase in a disordered
sample, or, as shown in Refs.~\cite{demler,sarma2}, a Bose glass of
the singlet bosons.

In summary, we have studied the double-layer electron system in a
parabolic quantum well at $\nu=2$ at tilted magnetic fields using
capacitance spectroscopy. We observe a change of the ground state of
the system at the Zeeman splitting far smaller than $\Delta_{SAS}$.
At the transition point the activation energy is found to be finite
although the ground state is composed of domains of two competing
states. Our data correspond well to insulator-insulator quantum phase
transition from a spin unpolarized to canted antiferromagnetic state
in a disordered system \cite{demler}.

We are thankful to V.~Pellegrini for valuable discussions. This work
was supported in part by Deutsche Forschungsgemeinschaft, AFOSR under
Grant No.~F49620-94-1-0158, the Russian Foundation for Basic Research
under Grants No.~97-02-16829 and No.~98-02-16632, and the Programme
"Nanostructures" from the Russian Ministry of Sciences under Grant
No.~97-1024. The Munich - Santa Barbara collaboration has also been
supported by a joint NSF-European Grant and the Max-Planck research
award.


\begin{references}
\bibitem{rem} Both the shift of the ground state energy and the
exchange-enhanced $g$ factor are caused by electron-electron
interaction. Nevertheless, these effects are quite different because
the former corresponds to $k=0$ whilst the latter describes
excitations with $k\gg 1/l$. This difference was not clearly
expressed in Ref.~\cite{dol}.
\end{references}
\end{document}